\documentclass[]{article}
\usepackage{lmodern}
\usepackage{setspace}
\usepackage{amssymb,amsmath}
\usepackage{ifxetex,ifluatex}
\usepackage{fixltx2e} % provides \textsubscript
\ifnum 0\ifxetex 1\fi\ifluatex 1\fi=0 % if pdftex
  \usepackage[T1]{fontenc}
  \usepackage[utf8]{inputenc}
\else % if luatex or xelatex
  \ifxetex
    \usepackage{mathspec}
  \else
    \usepackage{fontspec}
  \fi
  \defaultfontfeatures{Ligatures=TeX,Scale=MatchLowercase}
\fi
\IfFileExists{upquote.sty}{\usepackage{upquote}}{}
\IfFileExists{microtype.sty}{%
\usepackage{microtype}
\UseMicrotypeSet[protrusion]{basicmath} % disable protrusion for tt fonts
}{}
\usepackage[margin=1in]{geometry}
\usepackage{hyperref}
\hypersetup{unicode=true,
            pdftitle={Data scraping, ingestation, and modeling: bringing data from cars.com into the intro stats class},
            pdfauthor={Sarah McDonald and Nicholas J. Horton},
            pdfborder={0 0 0},
            breaklinks=true}
\usepackage{color}
\usepackage{fancyvrb}

\DefineVerbatimEnvironment{Highlighting}{Verbatim}{commandchars=\\\{\}}
\usepackage{framed}
\definecolor{shadecolor}{RGB}{248,248,248}
\newenvironment{Shaded}{\begin{snugshade}}{\end{snugshade}}

\newcommand{\DataTypeTok}[1]{\textcolor[rgb]{0.13,0.29,0.53}{#1}}

\newcommand{\KeywordTok}[1]{\textcolor[rgb]{0.13,0.29,0.53}{\textbf{#1}}}
\newcommand{\NormalTok}[1]{#1}
\newcommand{\OperatorTok}[1]{\textcolor[rgb]{0.81,0.36,0.00}{\textbf{#1}}}

\newcommand{\StringTok}[1]{\textcolor[rgb]{0.31,0.60,0.02}{#1}}

\usepackage{graphicx,grffile}
\makeatletter
\def\maxwidth{\ifdim\Gin@nat@width>\linewidth\linewidth\else\Gin@nat@width\fi}
\def\maxheight{\ifdim\Gin@nat@height>\textheight\textheight\else\Gin@nat@height\fi}
\makeatother
\setkeys{Gin}{width=\maxwidth,height=\maxheight,keepaspectratio}
\IfFileExists{parskip.sty}{%
\usepackage{parskip}
}{% else
\setlength{\parindent}{0pt}
\setlength{\parskip}{6pt plus 2pt minus 1pt}
}
\providecommand{\tightlist}{%
  \setlength{\itemsep}{0pt}\setlength{\parskip}{0pt}}
\ifx\paragraph\undefined\else
\let\oldparagraph\paragraph
\renewcommand{\paragraph}[1]{\oldparagraph{#1}\mbox{}}
\fi
\ifx\subparagraph\undefined\else
\let\oldsubparagraph\subparagraph
\renewcommand{\subparagraph}[1]{\oldsubparagraph{#1}\mbox{}}
\fi

\let\rmarkdownfootnote\footnote%
\def\footnote{\protect\rmarkdownfootnote}

\usepackage{titling}

  \title{Data scraping, ingestation, and modeling: bringing data from cars.com
into the intro stats class}
    \pretitle{\vspace{\droptitle}\centering\huge}
  \posttitle{\par}
    \author{Sarah McDonald and Nicholas J. Horton}
    \preauthor{\centering\large\emph}
  \postauthor{\par}
      \predate{\centering\large\emph}
  \postdate{\par}
    \date{August 28, 2018}

\begin{document}
\maketitle

\hypertarget{introduction}{%
\subsection{Introduction}\label{introduction}}

New tools have made it much easier for students to develop skills to
work with interesting data sets as they begin to extract meaning from
data. To fully appreciate the statistical analysis cycle, students
benefit from repeated experiences collecting, ingesting, wrangling,
analyzing data and communicating results. How can we bring such
opportunities into the classroom? We describe a classroom activity,
originally developed by Danny Kaplan (Macalester College), in which
students can expand upon statistical problem solving by hand-scraping
data from cars.com, ingesting these data into R, then carrying out
analyses of the relationships between price, mileage, and model year for
a selected type of car.

Most students might be interested in car prices since many will be
purchasing a car at some point in the near future. This activity can
help them develop better understanding of factors associated with car
prices.

The revised GAISE (Guidelines for Assessment and Instruction in
Statistics Education) College report (2016) notes the importance of
multivariate thinking and the use of technology. Car prices, model year,
and mileage are all factors to consider when purchasing or selling a
car. Introductory statistics courses need to move beyond only addressing
bivariate questions to be able to explore multivariate relationships.

In an increasingly data-rich society, plenty of information is available
to prospective car purchasers. Consumers can analyze and compare
multiple cars to try to get the best deal. By gathering data by hand
from cars.com then using this information to generate multivariable
visualizations and model prices, students gain experience (1) working in
groups, (2) practicing undertaking reproducible analyses, and (3)
exploring a multivariate dataset. These key ideas of data generation,
data ingestion, data visualization for multivariate analyses, and data
modeling are reinforced throughout the activity.

We begin by describing the activity, sharing examples of data,
visualizations, and models, then suggesting possible extensions and
providing concluding thoughts. Instructor materials and datasets
associated with this activity can be found at
\url{https://github.com/Amherst-Statistics/Cars-Scraping-Webinar}.

\hypertarget{activity-class-one}{%
\subsection{Activity: Class One}\label{activity-class-one}}

Students work in pairs of two and use two computers to gather and
hand-enter data concerning the cost of a specific model of a car, then
analyze the variations in pricing, price associations with mileage and
age, the rate at which cars depreciate, and the cost of driving one
mile. One student reads off data from cars.com and the other enters the
data into a spreadsheet. Each pair is assigned a different city.

The first step of the activity involves gathering data from
\href{https://www.cars.com}{\emph{cars.com}}. Using the `advanced
filter' option, the model and make of the car are specified, along with
the assigned location and restriction to recent years. Various
components in price determination include the model, year, mileage, and
location.

\begin{figure}
\centering
\includegraphics{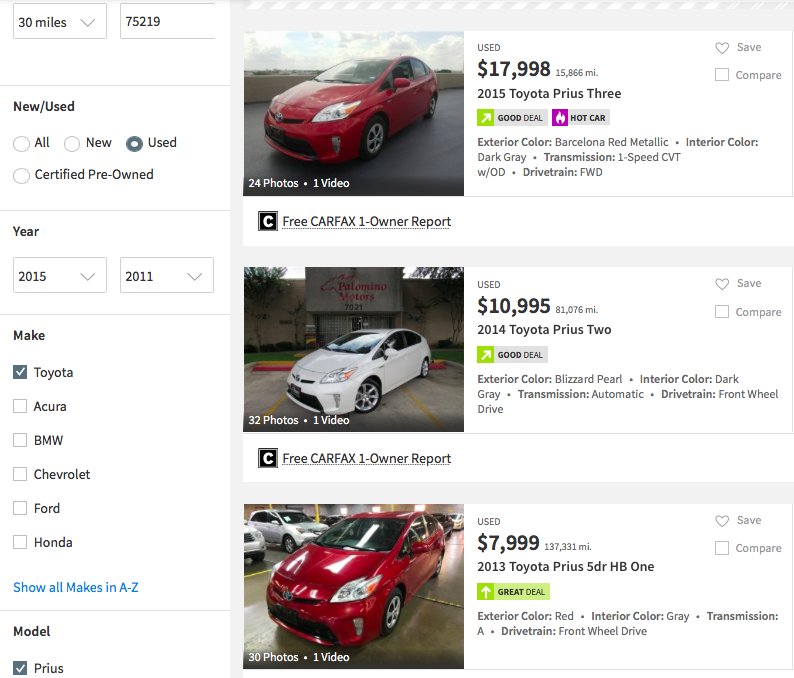}
\caption{Cars.com example snapshot of used Toyota Prius models from
Dallas}
\end{figure}

As an example, Figure 1 features a 2015 Toyota Prius from the Dallas
area, priced at \$17,998 with 15,866 miles whereas the 2014 Toyota Prius
is priced lower at \$10,995 but with a higher mileage of 81,076.
\newpage

Figure 2 illustrates data gathered and entered into an Excel sheet for a
group assigned to find car prices in Dallas.

\begin{figure}
\centering
\includegraphics{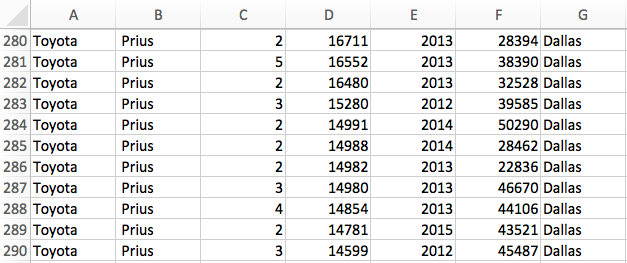}
\caption{Student hand-scraped data for Dallas entered into an Excel
spreadsheet}
\end{figure}

The data are entered into a spreadsheet (e.g., Excel, Open Office, or
Google Spreadsheet) using a template \texttt{cars.csv} to ensure that
the variable names are consistent between groups. Once the group has
completed the hand-scraping of 30 or 35 cars, they will upload this
spreadsheet into RStudio and run an instructor provided RMarkdown file
(\texttt{cars.Rmd}). The RMarkdown file reads the data that they have
uploaded to generate descriptive statistics, creates multivariate
displays, and fits a multiple regression model. The students need to
interpret the results and add their descriptions into the file.

The scatterplot produced in Figure 3 uses student-gathered data for
Toyota Prius to display the relationship between prices and mileage for
Dallas cars. The scatterplot reflects how car prices depreciate as a
function of mileage and model year. After the car's first year, the
discrepancy in price based on mileage by year tends to diminish.

The plot below displays a linear regression model for Prius prices in
Dallas.

\begin{figure}
\centering
\includegraphics{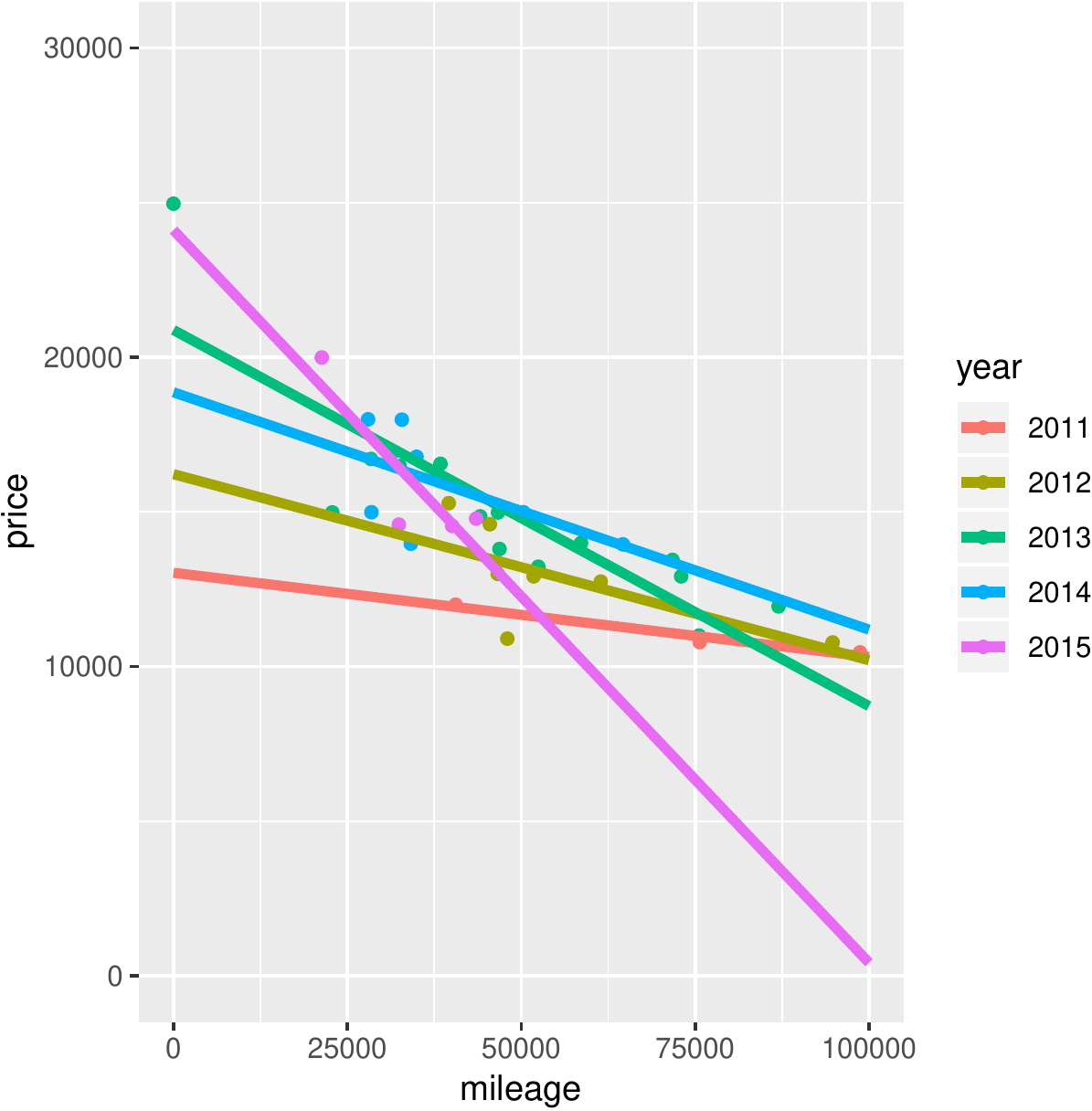}
\caption{Toyota Prius prices in Dallas based on mileage}
\end{figure}

Here the ggformula interface to the ggplot2 graphics system is used
because it provides a general modeling syntax similar to the `lm()'
function in R.

\begin{Shaded}
\begin{Highlighting}[]
\KeywordTok{library}\NormalTok{(ggformula)}
\KeywordTok{gf_point}\NormalTok{(price }\OperatorTok{~}\StringTok{ }\NormalTok{mileage, }\DataTypeTok{color =} \OperatorTok{~}\StringTok{ }\NormalTok{year, }\DataTypeTok{data =}\NormalTok{ Dallas) }\OperatorTok{%>%}
\StringTok{  }\KeywordTok{gf_lm}\NormalTok{()}
\end{Highlighting}
\end{Shaded}

\begin{table}[ht]
\centering
\begin{tabular}{rrrrr}
  \hline
 & Estimate & Std. Error & t value & Pr($>$$|$t$|$) \\ 
  \hline
(Intercept) & 19721.0125 & 706.8204 & 27.90 & 0.0000 \\ 
  mileage & -0.1075 & 0.0135 & -7.96 & 0.0000 \\ 
   \hline
\end{tabular}
\end{table}

The students then edit the RMarkdown file to interpret their results
based on the model and the graphical displays. For the Dallas group, the
summary output of the model in the table suggests that for every mile
driven, the car's predicted value (determined by price) will decrease on
average by about eleven cents.

Common errors that students experience include issues with formatting
(e.g., if they included dollar signs in the column for price) or
problems where they used different variable names than specified in the
assignment.

To obtain credit for the first part of the assignment, students must:

\begin{enumerate}
\def\labelenumi{\arabic{enumi})}
\tightlist
\item
  post the formatted file to RPubs (to allow a brief discussion of
  student findings and interpretations)
\item
  email the csv file to the instructor
\end{enumerate}

\hypertarget{activity-class-two}{%
\subsection{Activity: Class Two}\label{activity-class-two}}

Prior to the next class period, the instructor collates the data from
each group (in csv files) to create graphical displays, multiple
regression models, and interpretations from the data from all of the
cities. These results can be referenced as part of a future class
discussion. The collation process will identify issues (e.g.,
inconsistent formatting or variable naming) in the individual datasets,
which also provide an opportunity for discussion.

Figure 4 displays the scatterplot visualizing the relationship between
the price and mileage, where an interaction is included between the
mileage and (categorical) model year, using data scraped from all of the
cities (n = 830).

\begin{Shaded}
\begin{Highlighting}[]
\KeywordTok{library}\NormalTok{(mosaic)}
\KeywordTok{tally}\NormalTok{(}\OperatorTok{~}\StringTok{ }\NormalTok{location, }\DataTypeTok{data =}\NormalTok{ ds)}
\end{Highlighting}
\end{Shaded}

\begin{verbatim}
## location
##          40202        Atlanta     Bangor, ME    Baton Rouge         Boston 
##             40             40             40             40             40 
##        Buffalo        Chicago      Cleveland         Dallas    Los Angeles 
##             33             41             26             41             40 
##    Minneapolis    New Orleans            NYC        Phoenix       Portland 
##             59             33             40             39             40 
##       Richmond Salt Lake City      San Diego  San Francisco        Seattle 
##             40             33             39             39             39 
##          Tampa 
##             40
\end{verbatim}

We note that one group has included the zip code (needed to specify
location in cars.com) instead of the city name. Also note that some
groups only scraped 33 or 39 cars (to keep the class together on day one
data scraping was cut off after a certain amount of time).

\begin{figure}
\centering
\includegraphics{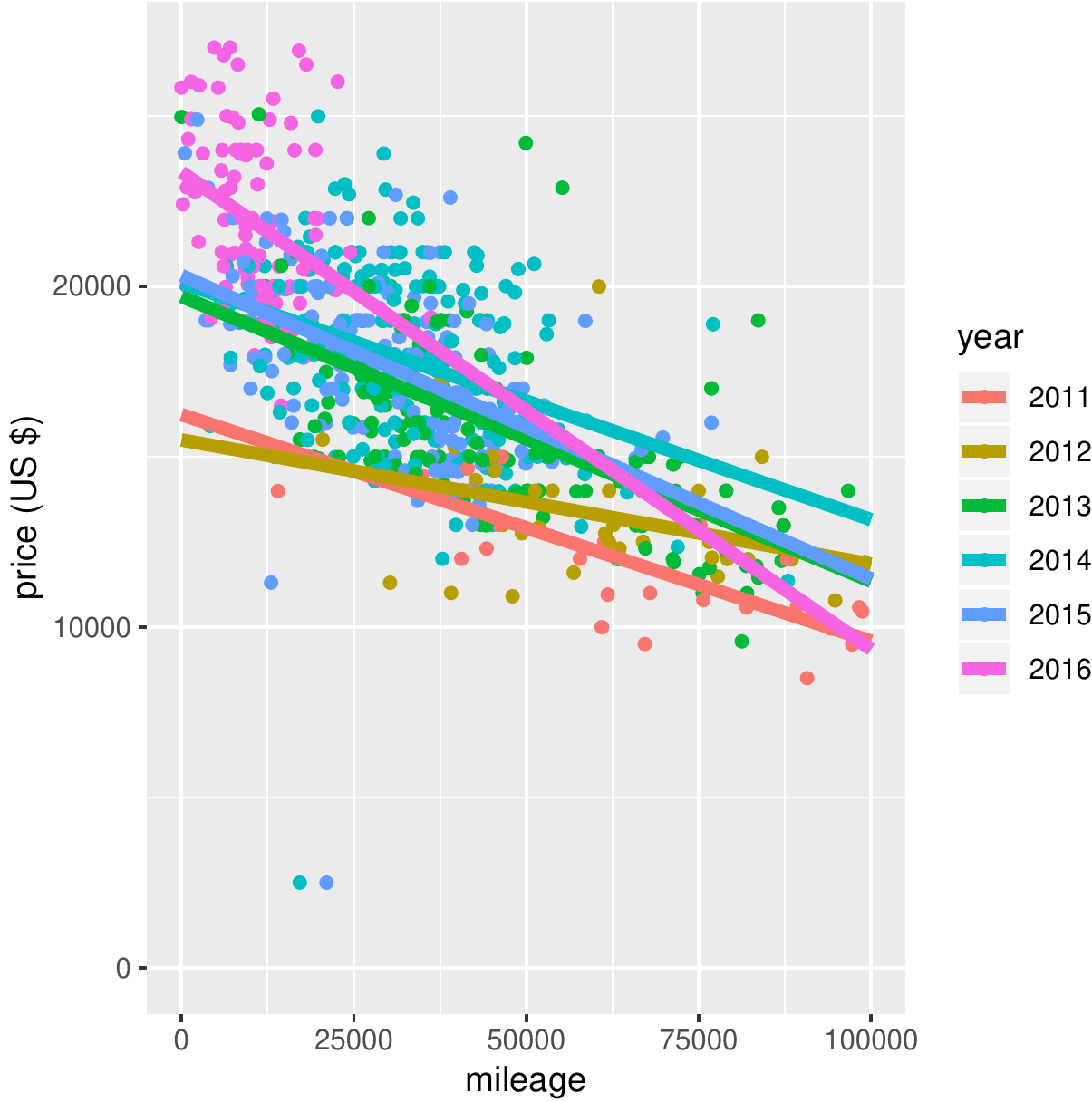}
\caption{Price versus mileage stratified by year (n = 830)}
\end{figure}

\begin{Shaded}
\begin{Highlighting}[]
\KeywordTok{gf_point}\NormalTok{(price }\OperatorTok{~}\StringTok{ }\NormalTok{mileage, }\DataTypeTok{color =} \OperatorTok{~}\StringTok{ }\NormalTok{year, }\DataTypeTok{data =}\NormalTok{ ds) }\OperatorTok{%>%}
\StringTok{  }\KeywordTok{gf_lm}\NormalTok{() }\OperatorTok{%>%}
\StringTok{  }\KeywordTok{gf_labs}\NormalTok{(}\DataTypeTok{y =} \StringTok{"price (US $)"}\NormalTok{)}
\end{Highlighting}
\end{Shaded}

\begin{table}[ht]
\centering
\begin{tabular}{rrrrr}
  \hline
 & Estimate & Std. Error & t value & Pr($>$$|$t$|$) \\ 
  \hline
(Intercept) & 17061.0694 & 868.8562 & 19.64 & 0.0000 \\ 
  locationAtlanta & -1638.4149 & 462.5576 & -3.54 & 0.0004 \\ 
  locationBangor, ME & -1689.6974 & 463.9047 & -3.64 & 0.0003 \\ 
  locationBaton Rouge & -745.2125 & 474.3208 & -1.57 & 0.1166 \\ 
  locationBoston & -563.6481 & 460.0693 & -1.23 & 0.2209 \\ 
  locationBuffalo & -581.6074 & 484.2352 & -1.20 & 0.2301 \\ 
  locationChicago & -2237.4990 & 456.4975 & -4.90 & 0.0000 \\ 
  locationCleveland & -1491.5866 & 520.8768 & -2.86 & 0.0043 \\ 
  locationDallas & -1078.1113 & 462.0475 & -2.33 & 0.0199 \\ 
  locationLos Angeles & 2319.6793 & 460.0475 & 5.04 & 0.0000 \\ 
  locationMinneapolis & -622.8958 & 423.7223 & -1.47 & 0.1419 \\ 
  locationNew Orleans & -573.2974 & 498.8439 & -1.15 & 0.2508 \\ 
  locationNYC & -594.5619 & 458.8934 & -1.30 & 0.1955 \\ 
  locationPhoenix & -325.9632 & 463.8124 & -0.70 & 0.4824 \\ 
  locationPortland & 65.2454 & 461.6683 & 0.14 & 0.8876 \\ 
  locationRichmond & -744.3217 & 461.1860 & -1.61 & 0.1069 \\ 
  locationSalt Lake City & -1954.0469 & 494.6800 & -3.95 & 0.0001 \\ 
  locationSan Diego & 257.6979 & 461.9773 & 0.56 & 0.5771 \\ 
  locationSan Francisco & 1578.2819 & 461.3929 & 3.42 & 0.0007 \\ 
  locationSeattle & 2136.5419 & 463.0608 & 4.61 & 0.0000 \\ 
  locationTampa & -2152.2974 & 462.1671 & -4.66 & 0.0000 \\ 
  mileage & -0.0606 & 0.0095 & -6.38 & 0.0000 \\ 
  year2012 & -251.3108 & 1135.1085 & -0.22 & 0.8248 \\ 
  year2013 & 3237.2317 & 894.6854 & 3.62 & 0.0003 \\ 
  year2014 & 3140.1907 & 888.3434 & 3.53 & 0.0004 \\ 
  year2015 & 3252.5139 & 885.3063 & 3.67 & 0.0003 \\ 
  year2016 & 8208.6105 & 874.4768 & 9.39 & 0.0000 \\ 
  mileage:year2012 & 0.0171 & 0.0144 & 1.19 & 0.2345 \\ 
  mileage:year2013 & -0.0180 & 0.0121 & -1.48 & 0.1394 \\ 
  mileage:year2014 & -0.0034 & 0.0140 & -0.25 & 0.8060 \\ 
  mileage:year2015 & -0.0099 & 0.0139 & -0.71 & 0.4778 \\ 
  mileage:year2016 & -0.1819 & 0.0275 & -6.60 & 0.0000 \\ 
   \hline
\end{tabular}
\end{table}

The multiple regression output describes the relationship between the
price based on location, mileage, year, and the interaction between
mileage and year. This is a relatively sophisticated model, with 32
predictors. Example interpretations of this model are included below:

LOCATION: After controlling for mileage and year, prices for a Toyota
Prius in Boston are predicted to be \$564 less than in Louisville,
Kentucky (the reference group). (Note the reference group is the first
group in the data set, which by R's default is alphabetically. Here, it
is Louisville, Kentucky as one group entered location as a zip code,
40202, rather than by name.)

MILEAGE: Holding location constant, the predicted price of a Prius
decreases on average by about six cents for an additional mile for
Priuses of the model.

INTERACTION: The interaction of mileage and year is more complicated to
interpret, since it includes five regression coefficients. We would
predict an additional average decrease in value of about eighteen cents
per mile driven for 2016 models compared with 2011 models, after
accounting for location. This is a great example of the \emph{new car
effect}: there is a much higher rate of depreciation in value of newer
cars in comparison to older models.

Other aspects of the model lend themselves to discussion. There are two
outliers (both from the same group) with very low prices. These are
likely prices that were entered incorrectly. In addition, the functional
form of the relationship between price and mileage (conditional on year)
is not very linear (though the regression model is assuming linear
relationships). We consider these as part of possible extensions of the
activity.

\hypertarget{extensions}{%
\subsection{Extensions}\label{extensions}}

In terms of introductory statistics, this activity works to develop
students ability to undertake the entire data analysis cycle. They
collect data by scraping information (by hand) from a website, then
loading this into RStudio.\\
With the data set, students can practice interpreting interaction terms
in the model. This practice will prove beneficial to students as data
sets (and models) become increasingly complex in future statistics
courses.

In the model produced in Figure 4, two outliers are observed. The two
points can be found in the data set by searching for Toyota Priuses
priced well below the average. Both data points indicate a pricing at
\$2,500 from Chicago, with one 2014 model and one 2015 model, and both
of the same model type (four). The 2014 model has a mileage of 17,152
wherein the average price for a used car of similar mileage in Chicago
is around \$15,550 and the 2015 model (with current mileage of 21,027)
would be priced around \$16,000, according to the model. It appears that
the large discrepancy between the price and mileage (well under the
average predicted price by \$13,000) could be due to input error, such
as a missing zero at the end of the value. Students should note these
outliers and decide from inference whether or not to include them in the
final model.

We have introduced this activity early in the course so have not focused
much on the functional form of the relationship between price and
mileage (beyond noting that the relationship is not very linear, see
Figure 5). Consideration of more flexible regression models could be
undertaken to better reflect the underlying relationships.

\begin{figure}
\centering
\includegraphics{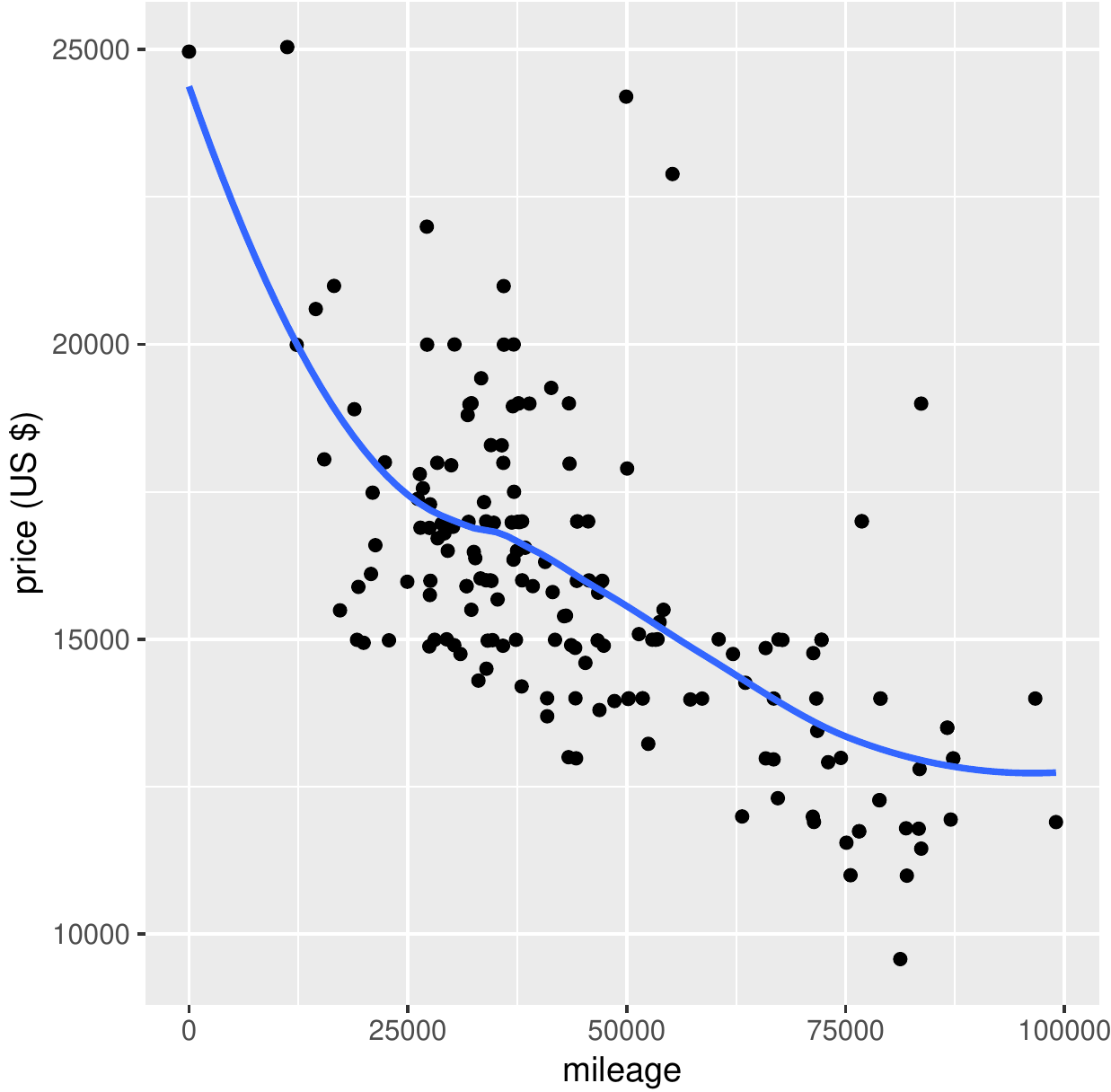}
\caption{Price versus mileage for 2013 vehicles (with superimposed
smoother)}
\end{figure}

While students included additional information in their spreadsheets
regarding trim models or add-on packages for the cars, this was not
incorporated into the modeling. Additional data wrangling would be
needed to bring this into the model as an additional predictor given the
inconsistent and idiosyncratic ways that such information is made
available by sellers in \emph{cars.com}.

Potential pitfalls include that the predictions made from the linear
models reflect only the cars in the data set and are not completely
representative of all car prices and locations. The models produced also
do not reflect consumer habits in its entirety as the data gathered only
demonstrates cars that are for sale and not necessarily sales price:
negotiation is important in determining sales price! Aspects of these
biases and data limitations could form the basis of a discussion of
design.

\hypertarget{conclusions}{%
\subsection{Conclusions}\label{conclusions}}

This activity is intended to reinforce critical aspects outlined by the
GAISE report, including teamwork, problem solving, and the use of data
to make decisions. This activity encourages multivariate thinking
through application facilitated by technology. The discovery of the
\emph{new car effect} is not obvious in a bivariate analysis.

Additional concepts such as data ingestion, regression modeling, and
graphical visualizations are among the other key learning outcomes.

Students are given the opportunity to gather data by hand and build
models to extract meaningful inferences. The learning objectives of the
cars.com activity permeate through other spheres of consumer habits and
students gain independence in their ability to make the best consumer
decisions. Financial literacy is an important capacity for students to
develop. This activity may help prepare students to make better
decisions when buying a car.

A focus on conceptual understanding, integration of real data with a
context and purpose, and a fostering of active learning are also
critical to students' comprehension. The usage of technology to explore
concepts and and analyze data, and assessments to improve and evaluate
student learning are additional goals of this activity.

\hypertarget{further-reading}{%
\subsection{Further Reading}\label{further-reading}}

GAISE College Report ASA Revision Committee, \emph{Guidelines for
Assessment and Instruction in Statistics Education College Report 2016,}
\url{http://www.amstat.org/education/gaise}.

National Academies of Sciences, Engineering, and Medicine. 2018.
\emph{Data Science for Undergraduates: Opportunities and Options.}
Washington, DC: The National Academies Press.
\url{https://doi.org/10.17226/25104},
\url{https://nas.edu/envisioningds}

Ben Baumer, Mine Cetinkaya-Rundel, Andrew Bray, Linda Loi, \& Nicholas
J. Horton (2014). RMarkdown: Integrating A Reproducible Analysis Tool
into Introductory Statistics. Technology Innovations in Statistics
Education, 8(1). Retrieved from
\url{https://escholarship.org/uc/item/90b2f5xh}

Randall Pruim, Daniel T. Kaplan, and Nicholas J. Horton. ``The Mosaic
Package: Helping Students to Think with Data Using R.'' R, June 2017,
journal.r-project.org/archive/2017/RJ-2017-024/.

Nicholas J. Horton, Benjamin S. Baumer, \& Hadley Wickham (2015) Taking
a Chance in the Classroom: Setting the Stage for Data Science:
Integration of Data Management Skills in Introductory and Second Courses
in Statistics, CHANCE, 28:2, 40-50, DOI: 10.1080/09332480.2015.1042739

\hypertarget{biographies}{%
\subsection{Biographies}\label{biographies}}

Sarah McDonald is a student at Amherst College, majoring in Statistics.
Her areas of interest include applications of statistical analysis in
consumer purchasing and behavioral habits. Her undergraduate research
involves studying effective ways to integrate and facilitate computation
in introductory statistics courses.

Nicholas J. Horton is Beitzel Professor of Technology and Society and
Professor of Statistics and Data Science at Amherst College, with
interests in longitudinal regression, missing data methods, statistical
computing, and statistical education. He received his doctorate in
biostatistics from the Harvard School of Public Health in 1999, and has
co-authored a series of books on statistical computing and data science.

\end{document}